\documentclass[graybox]{svmult}

\usepackage{mathptmx}       % selects Times Roman as basic font
\usepackage{helvet}         % selects Helvetica as sans-serif font
\usepackage{courier}        % selects Courier as typewriter font
\usepackage{type1cm}        % activate if the above 3 fonts are
\usepackage{makeidx}         % allows index generation
\usepackage{graphicx}        % standard LaTeX graphics tool
\usepackage{float}
\usepackage{multicol}        % used for the two-column index
\usepackage[bottom]{footmisc}% places footnotes at page bottom
\makeindex             % used for the subject index
\begin{document}

\title*{MHD equilibria and triggers for prominence eruption}
\titlerunning{Equilibria and Triggers for Eruption}
\author{Yuhong Fan}
\institute{Yuhong Fan \at National Center for Atmospheric Research,
3080 Center Green Dr., Boulder, CO 80301, \texttt{yfan@ucar.edu}.}
\maketitle

\abstract{Magneto-hydrodynamic (MHD) simulations of the emergence of twisted
magnetic flux tubes from the solar interior into the corona are discussed to
illustrate how twisted and sheared coronal magnetic structures (with free
magnetic energy), capable of driving filament eruptions, can form in the corona
in emerging active regions. Several basic mechanisms that can disrupt the
quasi-equilibrium coronal structures and trigger the release of the stored
free magnetic energy are discussed. These include both ideal processes such
as the onset of the helical kink instability and the torus instability of a
twisted coronal flux rope structure and the non-ideal process of the onset
of fast magnetic reconnections in current sheets. Representative MHD
simulations of the non-linear evolution involving these mechanisms are
presented.}

\section{Introduction}
\label{sec:intro}
Prominences/filaments are major precursors or source regions of coronal
mass ejections (CMEs) as indicated by the observed close association between
prominence/filament eruptions and CMEs
\cite{Munro:etal:1979,Webb:Hund:1987,Gopal:etal:2003}.
It is suggested that most CMEs are the
result of the destabilization and eruption of a prominence and its overlying
coronal structure, or of a magnetic structure capable of supporting a
prominence \cite{Webb:Hund:1987}.
On the large scale, prominences/filaments (either in strong active region
nests or in the weak fields of decaying active regions) represent stable
structures that can exist over long periods of time (days), that are much
longer than the Alfv\'en crossing time (minutes), before they
suddenly erupt. Thus, on the large scale for the prominence structure as a
whole, prominence magnetic fields represent magneto-static equilibria in the
corona.  Although on smaller scales, prominence plasma exhibits continuous,
on-going dynamic and turbulent behavior within the structure
(e.g. \cite{Berger:etal:2011, deToma:etal:2008, Liu:etal:2012} and see also
\cite{Chapter:Karpen}).

Given the low plasma $\beta$ in the
lower solar corona, where $\beta$ denotes the ratio of the plasma pressure over the
magnetic pressure, and if the gravitational force of the prominence mass is
not significant to distort the magnetic fields on which it resides, one class
of models considers prominence magnetic fields as approximately force free
equilibrium structures with field aligned electric current, i.e.
\begin{equation}
\nabla \times {\bf B} = \alpha {\bf B}
\label{eq:fff}
\end{equation}
where $\alpha$ is the torsion or twist parameter being constant along each
field line but generally different for different field lines.
The field aligned current, which manifests as the twist or shear of the
magnetic field represents the free magnetic energy (in excess of
of the potential field energy) stored in the equilibrium fields that can be
released to drive the eruption.
%%There are two main types of such force free field models, which are referred to
%%as the "sheared arcade" and "flux rope" models in the literature. The
%%boudary between the two types are rather blurred. The difference is not
%%qualitative, but mainly in the total amount of twist in the magnetic field
%%for the filament channel/coronal cavity, with the flux rope models
%%referring to the more twisted ones.

Another class of models consider the weight of the prominence mass
as playing a significant role in both the energy storage and release
of the prominence magnetic structures (e.g. \cite{Kippenhahn:Schlueter:1957,
Low:Smith:1993, Low:Zhang:2002,Fong:Low:Fan:2002,Low:Fong:Fan:2003,
Petrie:Low:2005, Zhang:Low:2004,Zhang:Low:2005}).
In these models a significant local dip or distortion in the field lines
can be created due to the weight of the prominence plasma, which produces
significant cross-field current and hence additional free magnetic energy.
A sudden removal of the prominence mass through some physical mechanisms
can therefore release the stored free magnetic energy and drive eruptions.

Due to the nearly frozen-in evolution of the large scale coronal magnetic
field in the highly conducting plasma of the solar atmosphere and corona,
the magnetic helicity, a physical quantity that measures the topological
complexity of the magnetic field (such as the linkage and/or twistness of
the field) is nearly conserved
(e.g. \cite{Berger:Field:1984, Berger:1984}).
In a volume $V$ with all magnetic flux
closed within the volume, the magnetic helicity is given by
$H = \int_V {\bf A} \times {\bf B} \, dV$, where ${\bf A}$ is the vector
potential of the magnetic field ${\bf B}$ in $V$, i.e.
${\bf B} = \nabla \times {\bf A}$, and it can be shown that
$H$ is invariant to any gauge transformation
of ${\bf A} \rightarrow { \bf A} + \nabla \chi $ with $\chi$ being an
arbitrary scalar function of position, and thus $H$ is a well
defined quantity.
For example, two linked, untwisted closed flux tubes with
fluxes $\Phi_1$ and $\Phi_2$ respectively, have a magnetic helicity
of $H=2 \Phi_1 \Phi_2$, and a uniformly twisted closed magnetic torus
with $T$ winds of field line rotation about the axis over
the length of the torus and with a total toroidal flux
of $\Phi$ has a magnetic helicity $H = T \Phi^2$
(\cite{Berger:Field:1984}).
For the solar corona, we do not generally have an isolated,
closed magnetic flux system and the magnetic flux is generally
threading through the photosphere.  Therefore
a relative magnetic helicity for the magnetic field
above the photospheric $z=0$ is defined (\cite{Berger:Field:1984})
\begin{equation}
H_r = \int_{z > 0} ({\bf A} + {\bf A}_p ) \cdot
({\bf B} - {\bf P}) \, dV,
\label{eq:Hr}
\end{equation}
where ${\bf B}$ is the magnetic field in the unbounded half
space above $z = 0$, ${\bf A}$ is the vector potential for
${\bf B}$, ${\bf P}$ is the reference potential field having
the same normal flux distribution as ${\bf B}$ on the $z = 0$
boundary, and ${\bf A}_p$ is the vector potential for ${\bf P}$.
The relative magnetic helicity $H_r$ is invariant with
respect to the gauges for ${\bf A}$ and ${\bf A}_p$, and is
thus a well-defined measure of the linkage or twistness of
the coronal magnetic field
(\cite{Berger:Field:1984,Demoulin:2007}).
The evolution of $H_r$ in the corona is given by
(e.g. \cite{Demoulin:2007}):
\begin{equation}
\frac{d H_r}{dt} = - 2 \int_S {\bf A}_p \times
( {\bf v} \times {\bf B} ) \cdot {\hat {\bf z}} \, dS
+ \left( \frac{d H_r}{dt} \right )_{\rm diss.},
\label{eq:dHdt}
\end{equation}
where the first term on the right-hand-side corresponds to
integration of helicity flux over the photospheric surface
and the 2nd term corresponds to dissipation of $H_r$ in the corona.  
In the above ${\bf A}_p$ is the uniquely determined
vector potential of the potential magnetic field
with the gauge conditions: ${\bf A}_p \cdot {\hat {\bf z}} = 0$ on
$S$, and $\nabla \cdot {\bf A}_p = 0 $ in the corona above
$S$. It is shown that the helicity dissipation (2nd term) is
negligible for the nearly frozen-in evolution of the large scale
corona even including magnetic reconnections during
flares (\cite{Berger:1984}). Such constraint of magnetic helicity
conservation is playing an important role in the energy storage and
ultimate eruption of the filament/prominence magnetic fields
as described by \cite{Zhang:Low:2005}.
The net helicity transported into the corona through the
photosphere (first term in the right-hand-side of eq.
\ref{eq:dHdt}) via flux emergence from the interior
cannot be flared away and therefore the free magnetic energy
cannot be completely dissipated down to the minimum energy
potential field level (\cite{Woltjer:1958,Zhang:Low:2005,Demoulin:2007}).
The observed hemispheric pattern of the chirality of filament channels
(see section 3.1 in \cite{Chapter:Engvold}) is directly related to the sign of magnetic
helicity contained in the filament channel magnetic fields, where
a dextral (sinistral) filament channel preferred in the northern (southern)
hemisphere contains dominantly negative (positive) helicity or left-handed
(right-handed) twist.
The hemispheric pattern of the helicity of filament channels has its
origin in the accumulation of the helicity in
emerging active regions (e.g. \cite{Zhang:Low:2005, Mackay:vanB:2005,
Yeates:etal:2008b} and see also \cite{Chapter:Mackay}), which are observed to also
show preferentially negative
(positive) twist in the northern (southern) hemisphere, and this 
sign preference does not change with the solar cycles. Such accumulation
of net helicity in each hemisphere is ultimately removed by the bodily
ejection of the filament/prominence magnetic fields as coronal mass ejections
(\cite{Zhang:Low:2005}).

The main questions to be addressed in this chapter are (1) how sheared or
twisted structures form in the corona as a result of magnetic flux
emergence? and (2) what are the mechanisms that lead to the sudden disruption
of the quasi-equilibrium coronal structures and an explosive release of the
free magnetic energy?
Question (1) is discussed in section \ref{sec:buildup} with focus on
understanding the formation of strongly twisted emerging active regions
that develop X-ray sigmoids and/or sigmoid shaped filaments.
Example MHD simulations of the emergence of a twisted flux tube from the
solar interior into the atmosphere and corona are shown to demonstrate how
helicity and free magnetic energy are transported into the corona.  In
section \ref{sec:triggers} the basic mechanisms that can trigger
the dynamic eruption of the twisted/sheared coronal magnetic structures are
discussed and several MHD simulations of the non-linear evolution involving
these mechanisms are presented.
It is argued that current sheet formation and magnetic reconnection are
playing an important role in all stages of the evolution of the
prominence/filament magnetic fields, before, during and after the eruption.

\section{Emergence of twisted magnetic fields and build up of free energy
and helicity in the corona}
\label{sec:buildup}
Observations suggest that flare productive active regions
are associated with the emergence of twisted magnetic
flux from the solar interior (see review by \cite{Schrijver:2009}).
Vector magnetic field observations of the photospheric layer of such
active regions show that the transverse magnetic field at the polarity
inversion lines (PILs) tend to be strongly
sheared, i.e. tends to be closer to being parallel to the PILs rather than
being perpendicular as expected for a potential field configuration
(e.g. review by \cite{Schrijver:2009}).
And sometimes the transverse magnetic field shows an ``inverse-polarity
configuration'' pointing from the negative polarity to the positive polarity,
indicating a concave upturning field configuration at the PILs
(e.g. \cite{Lites:2005,Canou:etal:2009,Okamoto:etal:2008}).
Continuous observations by SOT of Hinode studied by \cite{Okamoto:etal:2008}
have found a temporal evolution of the
transverse magnetic field at the PIL from a ``normal-polarity'' configuration
(pointing from the positive to negative polarity as expected for a convex
arcade loop field) to an ``inverse-polarity configuration'', which was
interpreted as the signature of the emergence of a helical flux rope through
the photosphere.
Flare productive active regions also often show velocity shear at the
PILs and rotating sunspots, indicative of transport of twist into the
solar corona (e.g. \cite{Brown:etal:2003,Zhang:etal:2008,Jiang:etal:2012}),
and develop sigmoid-shaped X-ray loops and sigmoid-shaped filaments
in the corona (e.g. \cite{Chae:etal:2001,Gibson:etal:2002}, see also Figure
9 in \cite{Chapter:Engvold}).

\begin{figure}[htb!]
\centering
\includegraphics[width=0.42\textwidth]{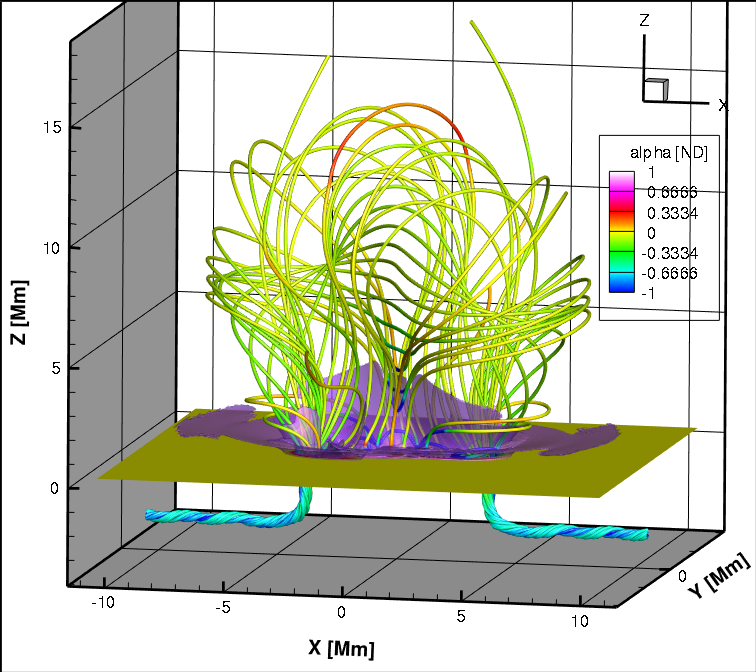}
\includegraphics[width=0.42\textwidth]{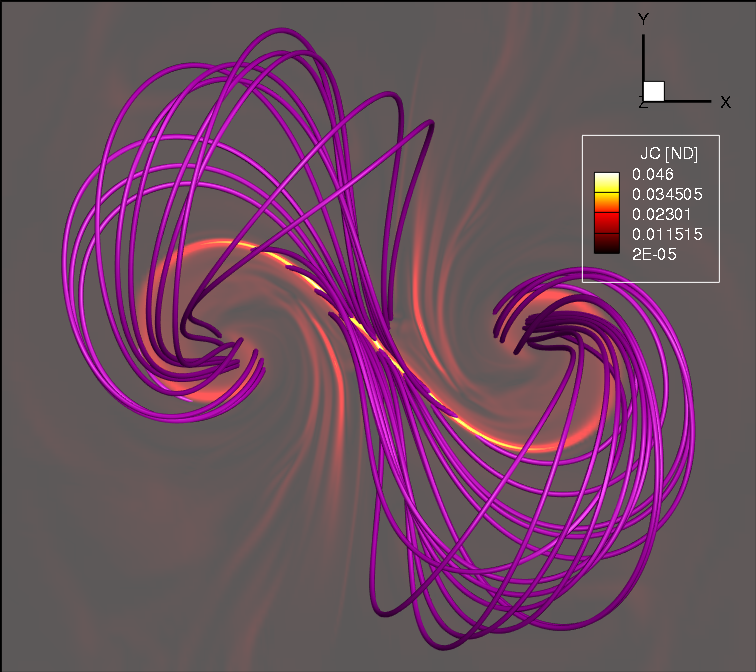} \\
\includegraphics[width=0.42\textwidth]{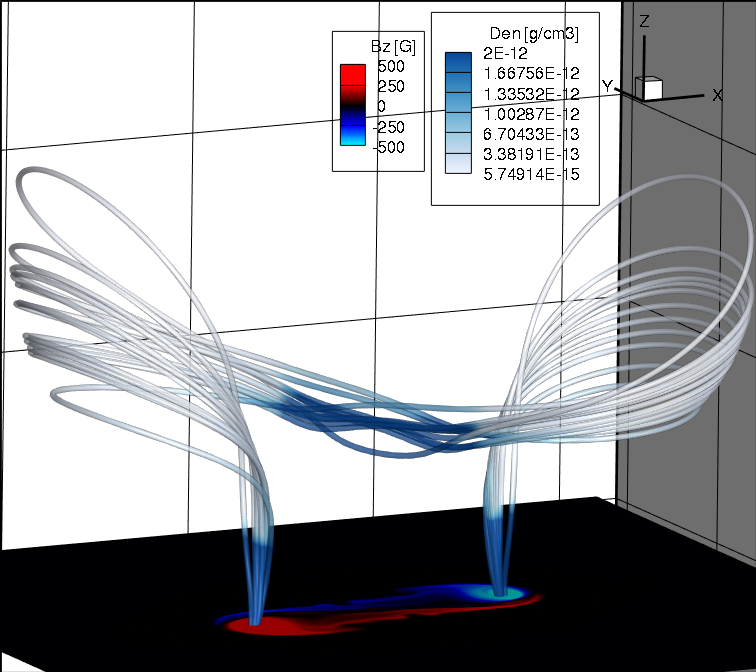}
\includegraphics[width=0.42\textwidth]{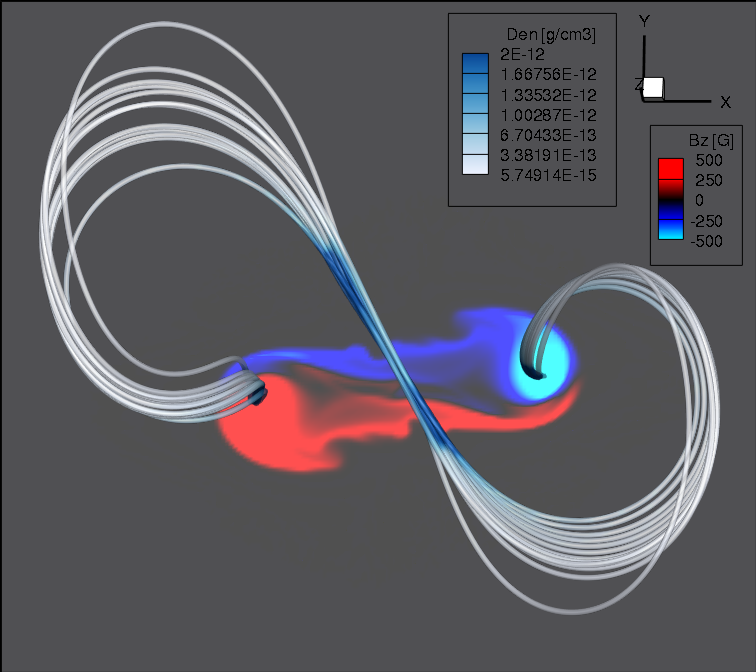} \\
\includegraphics[width=0.42\textwidth]{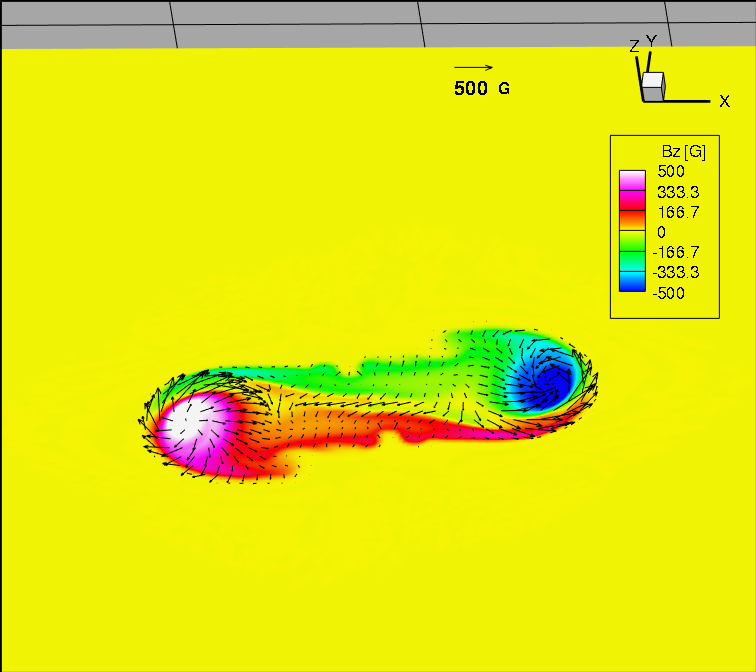}
\includegraphics[width=0.42\textwidth]{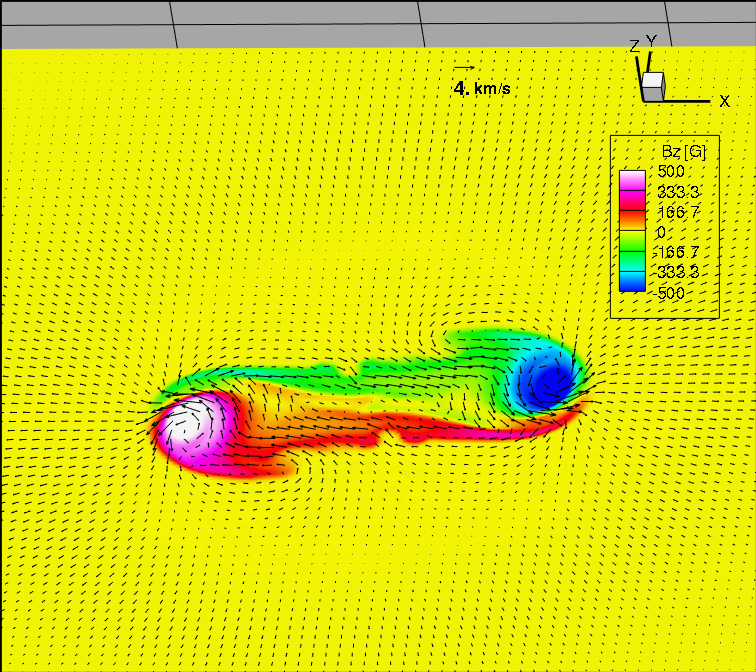}
%\picplace{5cm}{2cm} % Give the correct figure height and width in cm
\caption{Results from an example MHD simulation of
the dynamic emergence of a twisted subsurface flux tube into the solar
atmosphere and corona as described in \cite{Fan:2009}.
Top-left panel shows the 3D coronal
magnetic field structure with field lines colored by the torsion parameter
$\alpha$. The purple surface is the iso-surface of electric current density
$J$, outlining regions of high current concentration.
Top right panel shows sample field lines (in purple) going through the
central current sheet. The background color image shows a horizontal
cross-section of the current density $J$ (the color map is for $J$) at 3 Mm
above the photosphere. The two middle panels show two perspective views of
a set of field lines with central dips in the corona at a height about
5 Mm above the photosphere. The field lines are colored by the density,
showing enhanced density at the dips. The two bottom panels show the
vertical magnetic field pattern (color image) with arrows of the
transverse magnetic field (bottom left panel) and horizontal velocity
field (bottom right panel).}
\label{fig:fan2009}       % Give a unique label
\end{figure}

MHD simulations have shown that the emergence of a twisted magnetic flux
tube from the interior into the solar atmosphere and corona can qualitatively
explain many of these commonly observed features associated with strongly
flaring active regions (e.g. \cite{Magara:2004,Manchester:etal:2004,
Archontis:etal:2009, Fan:2009,Fang:etal:2010,Fang:etal:2012,
Fang:etal:2012b,Leake:etal:2013}).
Figure \ref{fig:fan2009} shows the results from one example simulation
(\cite{Fan:2009}) of the 3D coronal magnetic field structure
and the photospheric flux emergence patterns
produced by a subsurface twisted flux tube whose central segment rises
buoyantly to the photosphere and emerges into the atmosphere and the corona as
a result of the non-linear evolution of the magnetic buoyancy instability. 
It is found that the twisted subsurface flux tube does not emerge as a
whole into the corona. While the upper parts of the helical field lines
of the subsurface tube expand into the atmosphere due to the onset of
the magnetic buoyancy instability, the bottom U-shaped
portions of the winding field lines remain anchored at and below the
photosphere layer by the weight of the plasma. Nevertheless, the
simulations (e.g. \cite{Magara:2004,Manchester:etal:2004,
Archontis:etal:2009, Fan:2009, Leake:etal:2013}) show that a
flux rope structure with field lines winding about each other and with
sigmoid-shaped, dipped, core field lines eventually forms in the corona 
(top-left panel in Figure \ref{fig:fan2009}).
It is found that the Lorentz force drives both shear flows at the PIL
(\cite{Manchester:etal:2004}) and rotational motions in each of the
polarity concentrations (reminiscent of rotating sunspots) as shown in
the bottom right panel of Figure \ref{fig:fan2009}. 
Such vortical motions are caused by a gradient of the torsion or twist
$\alpha \equiv ((\nabla \times {\bf B}) \cdot {\bf B})/{\bf B}^2$ along
the field lines from the interior to the corona (see the $\alpha$ coloring
of the field lines in the top left panel of Figure \ref{fig:fan2009}) due to
the great expansion and stretching of the emerged coronal fields
(\cite{Longcope:Welsch:2000,Fan:2009}). The shear and vortical motions
are the major means by which twist or magnetic helicity are continually
transported from the interior flux rope into the corona in the
emerging region. \cite{Fan:2009} found that with the continued
twisting of the emerged field lines by the vortical motions at their
footpoints, the field lines above the PILs rotate and change their
orientation from an initial ``normal-polarity'' configuration into an
``inverse-polarity'' configuration (see bottom left panel of Figure
\ref{fig:fan2009}), leading to the formation of the sigmoid-shaped dipped
core fields (see the example field lines shown in the middle panels of
Figure \ref{fig:fan2009}).  This would explain the
observed rotation of the transverse field at the PIL described in
\cite{Okamoto:etal:2008}. With continued transport of twist into the
corona, the sigmoid-shaped core field also begins to rise 
upward into the corona, causing an underlying sigmoid-shaped vertical
current sheet to form (as outlined by the purple iso-surface of the current
density $J$ in the top left panel of Figure \ref{fig:fan2009}).
Reconnections in this vertical current sheet are of the ``tether-cutting'' type
(e.g. \cite{Moore:etal:2001}) that disengage the anchoring of the field lines
and allow the coronal flux rope structure to rise further in the corona (e.g.
\cite{Manchester:etal:2004,Fan:2009}, in some cases leading
to eruptive behavior (e.g. \cite{Archontis:Hood:2012, Archontis:etal:2014a}).
The top right panel of Figure \ref{fig:fan2009} shows the horizontal
cross-section of $J$ at 3 Mm above the photosphere showing the
sigmoid-shaped current concentration, and the sampled field lines
(in purple) going through the strong current concentration may
correspond to the observed sigmoid-shaped X-ray loops.

\begin{figure}[htb!]
\centering
\includegraphics[width=1.\textwidth]{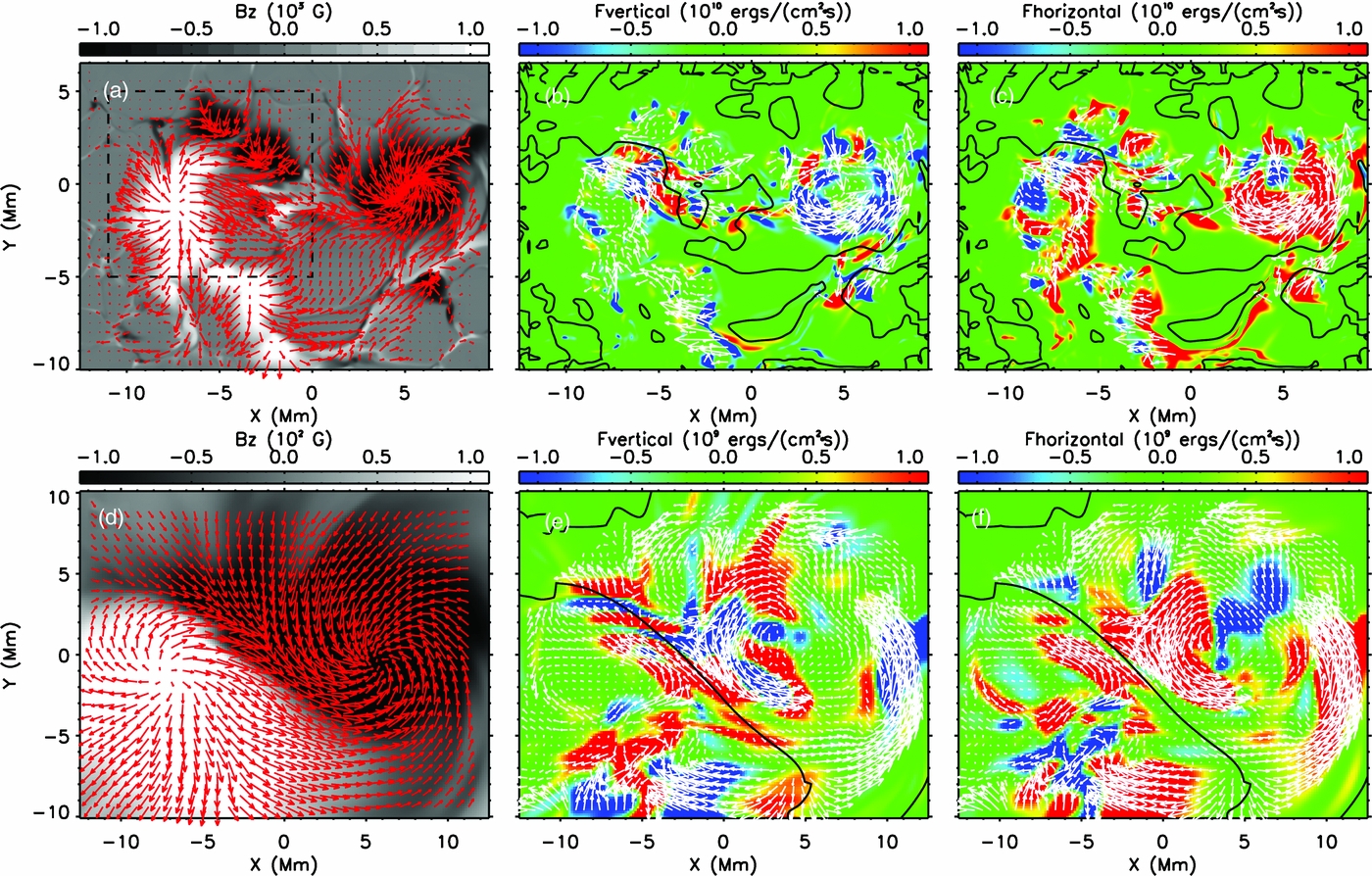}
\caption{Flux emergence pattern from a simulation of the emergence
of a twisted flux tube from the interior into the atmosphere and
corona including magneto-convection in the interior layer by
\cite{Fang:etal:2012b}: Grayscale image of
$B_z$ and red arrows of the transverse magnetic field
(a), Poynting flux $F_{\rm vertical}$ due to direct vertical motions
with white arrows of horizontal velocity fields (b), and Poynting
flux $F_{\rm horizontal}$ due to horizontal flow fields
with white arrows of horizontal
velocity (c), on the photosphere.  (d),(e), and (f) are
the same as (a), (b), and (c) respectively but at a height of
$z = 3$ Mm in the corona.  PIL is shown by the black line. One
clearly sees shear flows at the PIL, and a prominent rotation
of the negative polarity spot. Transport of magnetic energy into
the corona is clearly dominated by the component due to the
horizontal motions (shear and sunspot rotation). Figure from
\cite{Fang:etal:2012b} reproduced by permission of the AAS.}
\label{fig:fangetal2012}
\end{figure}

More recent simulations (\cite{Leake:etal:2013}) of twisted
flux tube emergence incorporating a pre-existing dipole coronal
field found clearly the formation of a stably confined coronal flux
rope structure with sigmoid shaped dipped core fields and an
underlying current sheet similar to that found in Figure \ref{fig:fan2009}.
They confirm the results that shear and sunspot
rotation are driven by twisted flux tube emergence and they can
cause the formation of stable sigmoids prior to a solar flare.
Simulations that incorporate magneto-convection
(\cite{Fang:etal:2012b}) in the interior
layer also found that shear flows at the PILs and sunspot rotation
driven by the Lorentz force are the major means twist and
free-magnetic energy are transported from the interior
into the corona (Figure \ref{fig:fangetal2012}).
It is found that sigmoid-shaped sheared fields are built up
in the corona, but the formation of a coherent flux rope structure
with dipped fields and inverse-polarity configuration has not
been seen in these simulations.

\section{Initiation mechanisms for eruption:}
\label{sec:triggers}
Due to the fast Alfv\'en speed in the lower solar corona ($\sim 1000$
${\rm km \, s^{-1}}$),
the process of magnetic flux emergence characterized by a photospheric
flow speed of order a few km/s represents a slow driving or change of
the coronal magnetic field.  Thus the resulting twisted coronal structure
(as discussed in the previous section) that forms is expected to evolve
quasi-statically through a sequence of near force free equilibria as
it is being driven slowly at the foot points by the continued shearing and
twisting produced by the flux emergence.
The reason that the quasi-equilibrium coronal structures suddenly erupt as
flares and/or coronal mass ejections is still under debate. The mechanism
that leads to the loss of a stable equilibrium and triggers the energy
release and eruption may be purely ideal or
involve non-ideal processes such as magnetic reconnections (e.g.
\cite{Forbes:etal:2006}).  One likely possibility is the onset of
an ideal-MHD instability or a sudden loss of an ideal-MHD equilibrium.  
Such mechanisms can naturally account for the fast Alfv\'en time scales
for the onset of the eruptions.

\subsection{Ideal MHD instabilities and loss of equilibrium of force free
coronal flux ropes}
A coronal magnetic field may suddenly erupt if the (force free) equilibrium
becomes unstable to perturbations, i.e. if the resulting forces produced
by the perturbations make the perturbations grow rather than restoring the
equilibrium, or if there is no more neighboring equilibria in the
evolution of the force free coronal magnetic field.
Both types of theoretical analysis of (1) the ideal linear instabilities of a
force free equilibrium (e.g.\cite{Hood:Priest:1981}, \cite{Kliem:Toeroek:2006},
\cite{Isenberg:Forbes:2007},\cite{Demoulin:Aulanier:2010})
or (2) the catastrophic loss of neighboring force free equilibrium solutions
beyond a certain value of some evolutionary parameter
(e.g. \cite{Forbes:Priest:1995}, \cite{Lin:etal:1998},
\cite{Demoulin:Aulanier:2010}) have been carried out.

Two current-driven instabilities that have been extensively studied as likely
triggers for flares and eruptions are the helical kink instability and the so
called ``torus instability'' associated with a twisted flux rope.
A force free cylindrically symmetric twisted flux tube of infinite length
is shown to be always unstable to the helical kink instability
(\cite{Anzer:1968}).
The kink instability can be suppressed if the ends of the cylindrical flux
tube are line-tied such that within the finite length of the flux
tube the total twist is not too high (\cite{Raadu:1972}).
Thus anchoring of the footpoints of the coronal loops by the heavy plasma
of the photosphere is stabilizing for the coronal magnetic field.
Again considering cylindrically symmetric force free magnetic flux tubes
line-tied at both ends, \cite{Hood:Priest:1981} show that for a uniformly
twisted flux tube the kink instability sets in if the angle each field line
is twisted in going from one end to the other exceeds 2.49 $\pi$, or about
1.25 full rotations. 

\begin{figure}[htb!]
\centering
\includegraphics[width=0.95\textwidth]{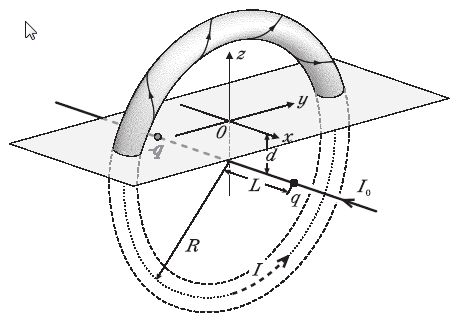}
\caption{The T\&D (Titov and Demoulin) force free coronal flux rope model from
\cite{Titov:Demoulin:1999}. Figure reproduced with permission by
Astron. \& Astrophys.}
\label{fig:td1999}
\end{figure}

On the other hand, the torus instability is an expansion instability
associated with a toroidal current ring held in equilibrium in an external
potential magnetic field (\cite{Kliem:Toeroek:2006},
\cite{Demoulin:Aulanier:2010}). The stability of such equilibrium
configurations has been studied in Tokamaks (e.g. \cite{Bateman:1978}).
The torus instability for an arched coronal flux rope confined
by an external potential field has been
demonstrated using an analytic model (\cite{Titov:Demoulin:1999},
\cite{Isenberg:Forbes:2007}) of an approximately force free coronal
magnetic field as shown in Figure \ref{fig:td1999}, hereafter referred to as the
T\&D (Titov and Demoulin) flux rope model.
The force free coronal magnetic field above the photosphere is constructed
using three sources: a circular flux rope with a thin total current $I$,
a pair of charges $q$ and $-q$ below the photosphere, and a line current
$I_0$ below the surface. The subsurface sources are just virtual sources
for the analytic construction of the normal magnetic flux at the photosphere,
and are not reality. With such construction, the forces at the apex of the
flux rope (\cite{Titov:Demoulin:1999}) can be
decomposed into an outward hoop force,
corresponding to the self-repulsive force of the circular current $I$,
and an inward confining force acting on the flux rope current due to the
potential field $B_p$ produced by the charges. Note, the subsurface line
current $I_0$ is introduced as a means of controlling the pitch of the
magnetic field in the vicinity of the coronal current and it does not
exert a force on the coronal current $I$ because the field $I_0$ generates
is parallel to $I$. In equilibrium the outward hoop force and the inward
confining force due to the potential magnetic field should
balance. \cite{Titov:Demoulin:1999} considered a sequence of equilibrium
states with increasing major radius $R$, which could be viewed as a
quasi-static emergence of the flux rope.  They found that the equilibrium
becomes unstable to an expansion $\delta R$, when $R$
reaches a critical size with respect to the separation of the charges, where
the decline of the potential field $B_p$ with $R$ becomes sufficiently steep,
as measured by a decay index of $n \equiv - d \ln B_p / d \ln R > 1.5$,
such that the decline of the confining force by the potential magnetic field
becomes faster than the decline of the outward hoop force. 
\cite{Titov:Demoulin:1999}'s calculation of the instability considered
an azimuthally symmetric expansion $\delta R$, and therefore does not
enforce line-tying by the heavy photosphere during the time scale for the
onset of the instability.  Later improved calculation of the torus instability
for the T\&D flux rope equilibrium by \cite{Isenberg:Forbes:2007}
considered perturbations that truly enforce anchoring of the coronal flux
rope on the photosphere by using a subsurface image current to wipe out
the change on the photosphere magnetic field produced by the change of the
coronal current.
They found similar results, that there is a critical height for
the apex of the coronal flux rope, above which stable equilibrium of the
flux rope confinement is not possible.

\cite{Demoulin:Aulanier:2010} examined both the loss of equilibrium and
the torus instability of coronal flux ropes with concentrated thin current
channels in the corona of either a straight or circular shape.
They found that a global instability of the magnetic configuration is present
when the current channel is located at a coronal height, $h$, large enough
so that the decay index of the potential field, $n \equiv -d \ln B_p/ d \ln h$,
is larger than a critical value that is in the range of 1 to 1.5.
They found that when a loss of equilibrium occurs the magnetic
configuration is also ideally unstable to the torus instability.

\subsection{MHD simulations of the eruption of coronal flux ropes}

The above analytical studies of the ideal linear instabilities and loss
of equilibrium of coronal force free magnetic fields are necessarily limited
to highly idealized field configurations with a high degree of symmetry.
To study realistic three-dimensional coronal magnetic field configurations
and the non-linear evolution of the instabilities and loss of equilibrium,
MHD simulations are important tools to provide physical insight.

\begin{figure}[htb!]
\centering
\includegraphics[width=0.95\textwidth]{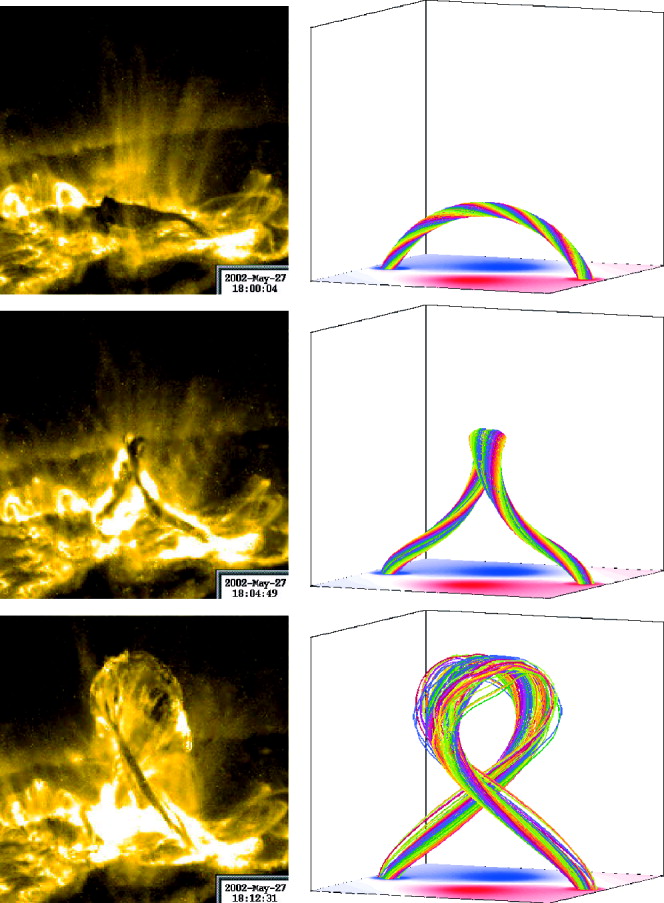}
\caption{{\it Left}: TRACE 195 {\AA} images of the confined filament eruption on
2002 May 27. {\it Right}: Magnetic field lines outlining the core of the
kink-unstable flux rope from the simulation of \cite{Toeroek:Kliem:2005}.
Figure from \cite{Toeroek:Kliem:2005} reproduced by permission of the AAS.}
\label{fig:tk2005}
\end{figure}

\cite{Toeroek:etal:2004, Toeroek:Kliem:2005} have performed detailed studies
of the helical kink instability of an arched coronal flux rope, line-tied to
the photosphere, using the T\&D analytical flux rope model as the initial
state in three-dimensional ideal MHD simulations. They have shown that
this model relaxes to a numerical equilibrium very close to the initial
analytical model in the case of subcritical twist and that the helical kink
instability develops for supercritical twist in the anchored flux rope.
It is found that the non-linear development of the kink instability with
supercritically twisted initial flux ropes can lead to either confined
(failed) eruptions or CME like ejective eruptions depending on
how rapidly the overlying field above the flux rope decreases with height.
By using a certain parameter setup of the T\&D flux rope with a supercritical
initial twist, \cite{Toeroek:Kliem:2005} is able to reproduce both the
development of the helical shape and the rise profile of a confined (or failed)
filament eruption observed on 27 May 2002 (see Figure \ref{fig:tk2005}).
The eruption of the kinked flux rope is halted by the strong overlying field
and reconnection outflows in the current sheet above the flux rope causes
expansion of the top part of the flux rope in the lateral directions as seen
in both the simulation and the observation.
A strong vertical current sheet also develops under the kinked flux rope
between the two legs of the kinked loop (e.g. \cite{Fan:Gibson:2004,
Toeroek:Kliem:2005}), which is consistent with the
coronal hard X-ray emission near the crossing of the loop legs
observed during the flare (\cite{Liu:Alexander:2009}).
The similarities between the helical shape of erupting
filament/prominence in many observed events and the magnetic field
morphology produced by MHD simulations of the evolution of kink unstable
coronal flux ropes (e.g. \cite{Toeroek:Kliem:2005,Fan:Gibson:2004,Gibson:Fan:2006})
suggest that the onset of the kink instability 
is a viable initiation mechanism for triggering many of such events, and that
flux ropes with substantial twist can exist or form in the solar corona
prior to eruption.

Further using initial T\&D flux rope configurations in the parameter regime
where the flux rope is subcritical for the onset of the helical kink
instability but supercritical for the expansion torus instability,  
\cite{Toeroek:Kliem:2007} studied the non-linear evolution of the torus
instability of line-tied coronal flux rope embedded in a potential field
with varying decline profiles with height.
It is found that the critical decay index $n_{\rm cr}$ of the potential
field for the torus instability of the line-tied 3D flux rope is similar
to the analytical result for a freely expanding toroidal current ring
(\cite{Kliem:Toeroek:2006}), with $n_{\rm cr} \sim 1.5$.  It is also found
that the acceleration profile for the eruption of the unstable flux rope 
depends on the steepness of the field decrease, corresponding to fast CMEs
for rapid decrease (as typical of compact active regions) and to slow CMEs
for gradual decrease (as typical of quiescent filament eruptions).

The above MHD simulations of the helical kink instability and torus
instability of coronal flux ropes have all started out with initial
configurations (T\&D analytical models) that are already supercritical
for the onset of the instabilities.  For studying how
unstable configurations come to being and the transition
from the quasi-static buildup phase to the dynamic eruptive phase,
there have been many MHD simulations of the buildup and eruption of 
coronal flux ropes driven at the lower boundary by various flux transport
processes including:
a slow imposed flux emergence (e.g. \cite{Amari:etal:2004,
Fan:Gibson:2007, Fan:2010, Fan:2012, Chatterjee:Fan:2013},
shearing and twisting motions in conjunction with flux cancelation at the PIL
due to photospheric diffusion (e.g. \cite{Amari:etal:2003a, Amari:etal:2003b,
Aulanier:etal:2010}).
The last flux transport process is important for the 
the formation of quiescent filament channels in decaying active regions (see
\cite{Chapter:Mackay}).
Dynamic MHD simulations of the emergence of a twisted flux tube from the solar
convection zone through the photosphere into the solar atmosphere 
as described in Section \ref{sec:buildup},
have shown that shearing motions at the PIL and twisting motions of sunspots
are naturally driven during the flux emergence, which transport twist from
the interior into the corona.
Furthermore, a vertical current sheet tends to develop underlying the emerged
field, and the associated ``tether-cutting'' reconnections in the current
sheet contribute to the buildup of the coronal flux rope and allow it to rise
into the corona (Section \ref{sec:buildup}).
This picture of current sheet formation and magnetic reconnections
contributing to the buildup of a coronal flux rope during the quasi-static
phase has also been found in coronal MHD simulations of flux rope eruption
(e.g. \cite{Aulanier:etal:2010, Fan:2010, Fan:2012}).
\cite{Aulanier:etal:2010} performed a coronal MHD simulation of an initially
potential, asymmetric bipolar field, which evolves as it is driven at the
lower boundary by sub-Alfv\'enic, line-tied shearing motions and a slow
magnetic diffusion that causes flux cancellation at the PIL.
It is found that flux cancellation at the PIL transforms sheared arcades
into a slowly rising and stable flux rope. Later a quasi-separatrix  
layer (QSL see more discussion below) topology develops with the formation of
a vertical current sheet along
the QSL and ``tether-cutting'' reconnections allow the flux rope to
continue to rise slowly.  As the flux rope reaches the height at which the
decay index of the potential magnetic field reaches about 1.5, the flux
rope undergoes a rapid acceleration.  The conclusion from this study is that
the non-ideal resistive processes of photospheric flux cancellation and
tether-cutting coronal reconnections are not the trigger of eruption but are
the key pre-eruptive mechanisms for the build up and rise of the coronal
flux rope to the critical height at which the ideal torus instability
causes the eruption.

\begin{figure}[htb!]
\centering
\includegraphics[width=0.3\textwidth]{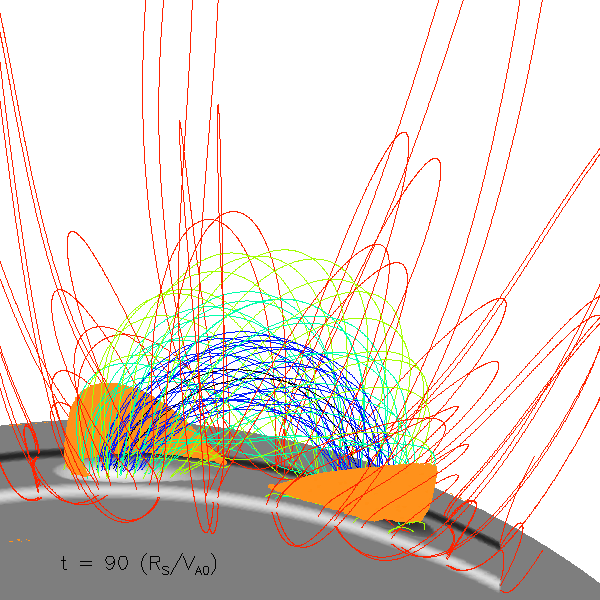}
\includegraphics[width=0.3\textwidth]{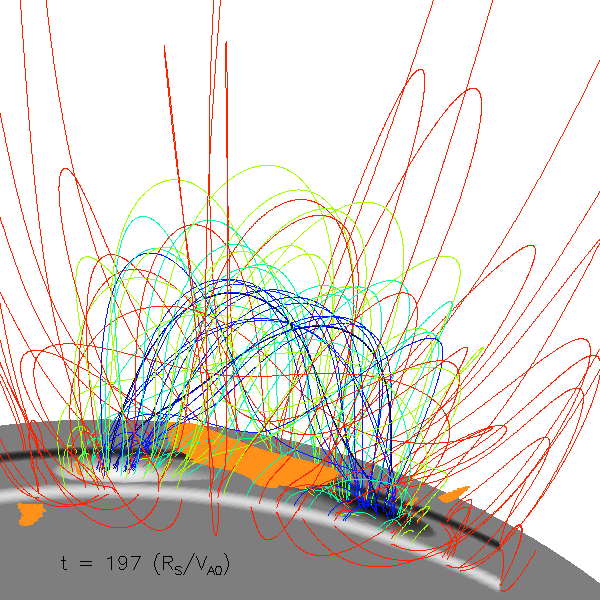}
\includegraphics[width=0.3\textwidth]{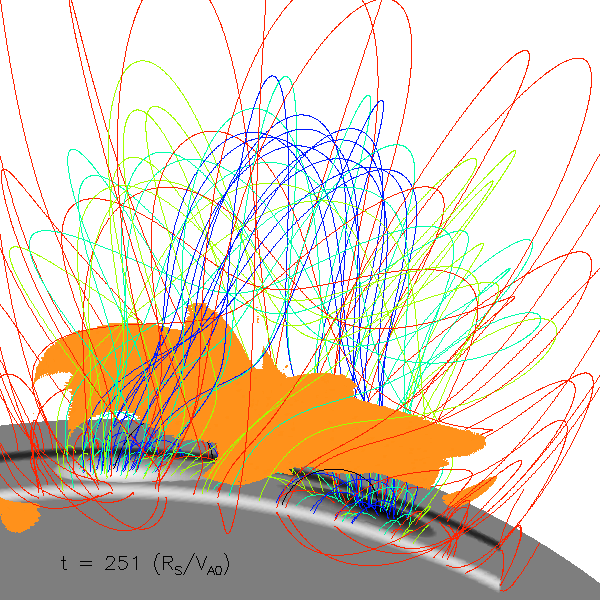} \\
\includegraphics[width=0.6\textwidth]{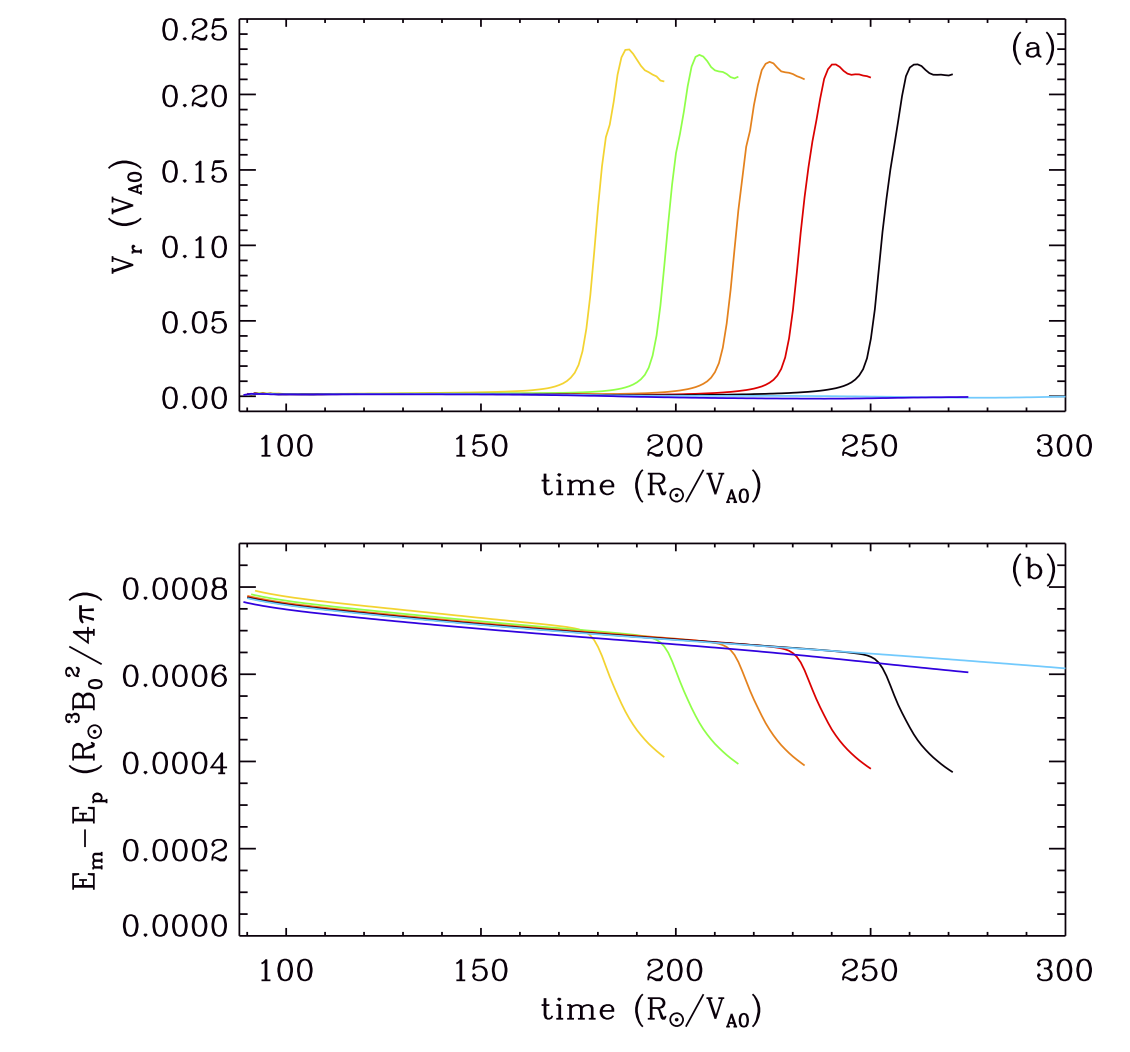}
\includegraphics[width=0.55\textwidth]{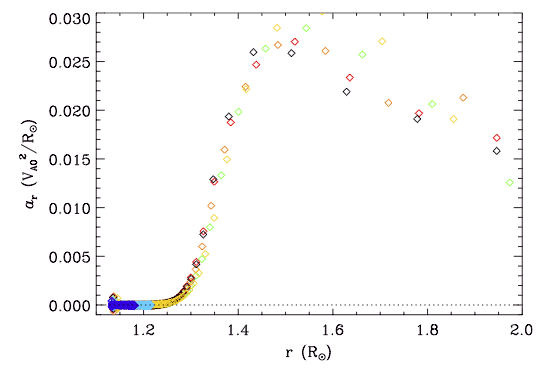}
\caption{Simulations by \cite{Fan:2010} of the buildup and eruption of
torus-unstable coronal flux ropes. Top row panels: snapshots of 3D
coronal flux rope with underlying current sheet formation
(orange iso-surfaces)
during the quasi-static rise phase (two left panels) and at the onset
of dynamic eruption when the critical height for the onset of the torus
instability is reached (right panel).
The evolution corresponds to the black curves in the lower panels.
2nd and 3rd-row panels show respectively the
flux rope rise speed and the free magnetic energy as a function of time
after the emergence is stopped, for the
different cases (different colored curves) where different amount of
twist is transported into the corona during the imposed flux emergence.
Bottom panel: the acceleration of the flux rope as a function of
the apex height for the different cases. Rapid acceleration takes place
when roughly the same critical height is reached. Figure reproduced by
permision of the AAS.}
\label{fig:fan2010}
\end{figure}

\begin{figure}[htb!]
\centering
\includegraphics[width=1.\textwidth]{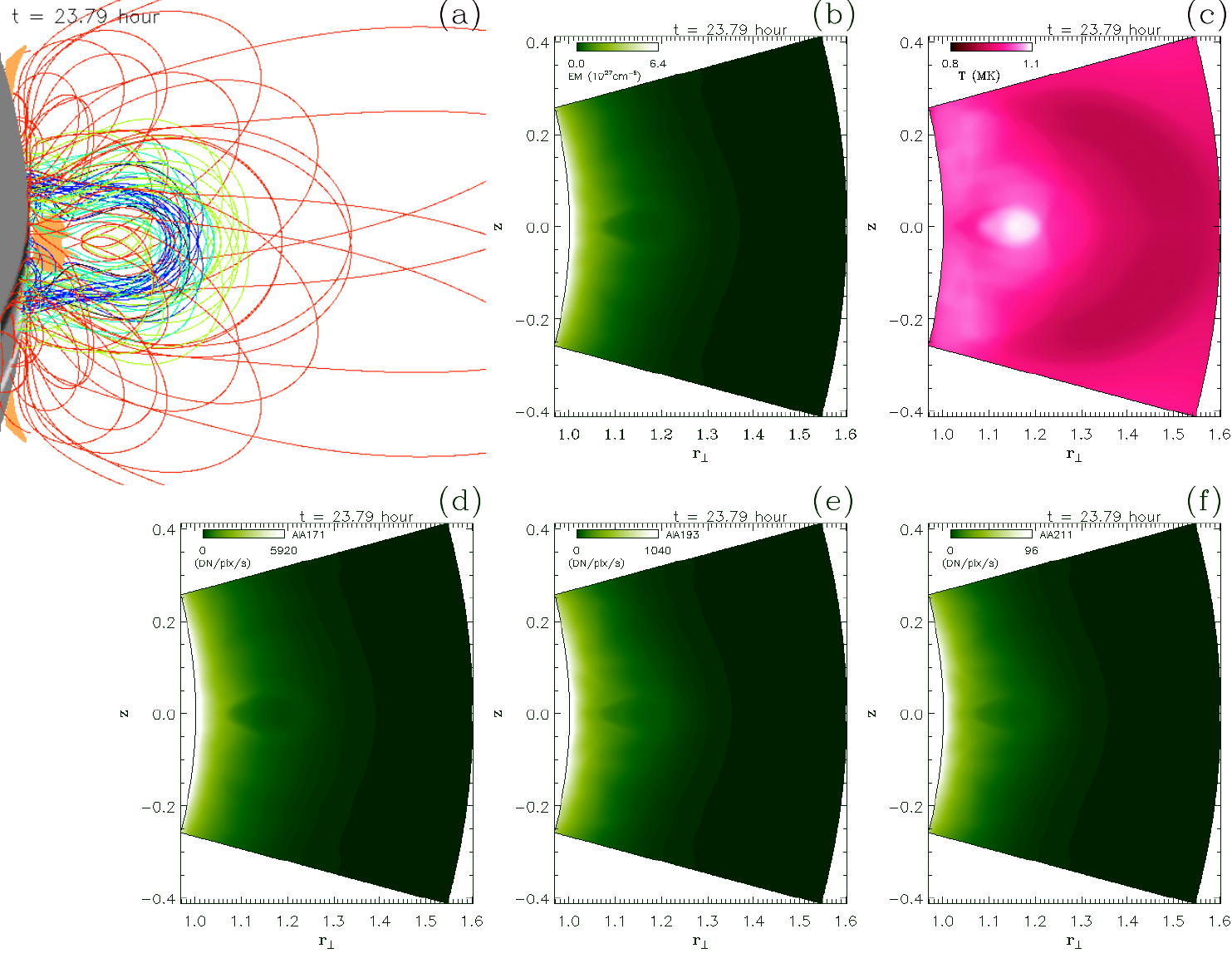}
\caption{From \cite{Fan:2012}. (a) 3D flux rope field lines and the
underlying current sheet (orange iso-surface of $J/B$) viewed above
the limb along the line of sight that is roughly aligned with the hot
channel. The hot channel forms due to the accumulation of
the reconnected flux rising into the flux rope produced by the
``tether-cutting'' reconnections during the quasi-static rise phase.
(b) and (c) show respectively the modeled emission measure and line-of-sight
averaged temperature showing the central hot low density cavity caused by the
hot channel, and (d), (e), and (f) show respectively the synthetic SDO/AIA
intensity images at 171 {\AA}, 193 {\AA}, and 211 {\AA}, as viewed from the
same line of sight. Figure from \cite{Fan:2012} reproduced by permission
of the AAS.}
\label{fig:fan2012}
\end{figure}

Similar evolution for the build up of torus-unstable coronal
flux ropes is found in the simulations of \cite{Fan:2010, Fan:2012}.
In \cite{Fan:2010} a set of simulations are carried out where the
coronal evolution is driven at the lower boundary by the
slow emergence of a twisted flux rope into a pre-existing coronal arcade
field, with a varying amount of twist being transported into the corona
for the different cases before the emergence is stopped.
In all the cases the emerged flux rope settles into a quasi-equilibrium
after the emergence is stopped.
Subsequently, the flux rope continues to rise quasi-statically due to
the ``tether-cutting'' reconnection in the vertical current sheet that forms
underlying the flux rope (the two left panels in the top row of Figure \ref{fig:fan2010}),
even though no more Poynting flux or helicity flux is transported into
the corona and the coronal magnetic energy is slowly
declining (3rd row panel in Figure \ref{fig:fan2010}). 
The flux rope is found to transition to a dynamic eruption (top right panel
in Figure \ref{fig:fan2010}) with rapid
acceleration and sharp release of the magnetic energy at varying times
for the different cases (the 2nd and 3rd-row panels of Figure \ref{fig:fan2010}),
but all corresponding to when roughly the same critical height is reached
(bottom panel of Figure \ref{fig:fan2010}) where
the decay index of the potential field reaches about 1.7.
This nearly uniform height dependence for the onset of eruption suggests
that the trigger of the eruption is caused by the onset of the torus
instability, while the tether-cutting reconnection is contributing
to the build-up of the flux rope and facilitating its (quasi-static) rise
to the critical height for the onset of the instability.
\cite{Fan:2012} further showed that the thermal signatures of such
``tether-cutting'' reconnections is the formation of a hot central low-density
channel containing reconnected, twisted flux threading under the flux rope
axis and on top of the central vertical current sheet.
When viewed in the line of sight roughly aligned with the hot channel
(see Figure \ref{fig:fan2012}), the central vertical current sheet
appears as a high-density vertical column with upward extensions as a
U-shaped dense shell enclosing a central hot, low-density void corresponding to
the central hot channel. Such
thermal features may correspond to the observed sub-structures (central hot
cavity on top of prominence ``horns'') that have been observed within coronal
prominence cavities (see \cite{Chapter:Gibson}).

\begin{figure}[htb!]
\centering
\includegraphics[width=0.99\textwidth]{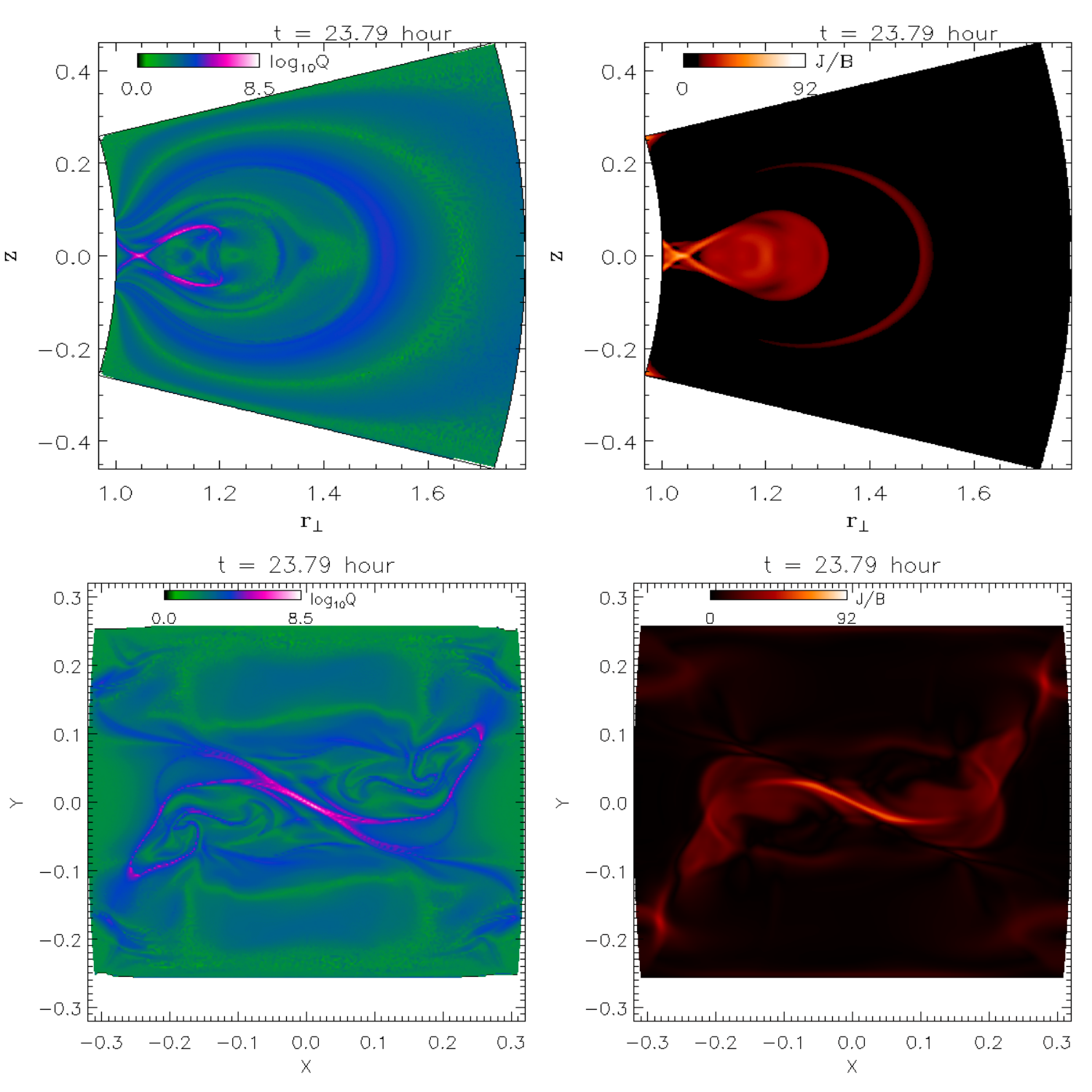}
\caption{From the simulation of \cite{Fan:2012}.
Squashing degree $Q$ in respectively the central meridional cross-section
of the flux rope (upper left panel) and a horizontal cross-section 
below the hot channel of the flux rope (lower left panel).  The high
Q value outlines the location of the QSLs, along which intense current
sheets form as shown in the corresponding cross-sections of $J/B$
in the right panels. Figure from \cite{Fan:2012} reproduced by permission
of the AAS.}
\label{fig:fan2012b}
\end{figure}

The central vertical current sheet underlying the simulated flux rope (Figure \ref{fig:fan2012}(a))
is found to have formed along topological structures
identified as quasi separatrix layers (QSLs, see Figure \ref{fig:fan2012b}),
which are regions of the magnetic volume where the field line connectivity
to the line-tied surface experiences drastic variations
(\cite{Demoulin:etal:1996a, Demoulin:etal:1996b,Titov:etal:2002}).
They are identified by estimating the so-called squashing degree, $Q$,
which measures the ``squashing'' of an elementary flux tube cross-section as
it is mapped from one foot point to the other (\cite{Titov:etal:2002,
Titov:2007, Pariat:Demoulin:2012}). QSLs correspond to regions of very large Q.
They are a generalization of the concept of separatrices at which the field
line linkage is discontinuous.
Similar to a separatrix, a QSL divides the coronal domain into
quasi-connectivity domains, and due to the drastic change of the field
line connectivity at the QSL, it is a site along which
current sheets or magnetic tangential discontinuities tend to form and
where magnetic reconnections take place (e.g. \cite{Demoulin:etal:1996b,
Aulanier:etal:2005,Savcheva:etal:2012}).
The QSL with the highest Q values shown in the central meridional cross-section
of the simulated flux rope (top left panel of Figure \ref{fig:fan2012b}) correspond to
the mid cross-section of the
so-called Hyperbolic Flux Tube (HFT), a generalization of the X-line
configuration, which divides the magnetic volume into four distinct domains
of magnetic field connectivity (e.g. \cite{Titov:2007, Aulanier:etal:2005,
Savcheva:etal:2012}). The central vertical current sheet underlying the flux rope
forms along the HFT (see the top right panel of Figure \ref{fig:fan2012b}).
\cite{Fan:2012} suggests that the observed density feature of a dense prominence
column with upward extending ``horns'' (see \cite{Chapter:Gibson}) corresponds to the
current sheet that forms along the HFT.

As has been discussed above, magnetic reconnection may be playing 
an important role for the build up of an unstable coronal flux rope before
eruption.  It has also been shown that magnetic reconnection is  
critically important for producing an ejective eruption of the magnetic
flux in a CME. The energetics for the magnetic storage models of CMEs
requires that the magnetic energy of the pre-eruption field exceeds the
subsequent magnetic energy during and after the eruption.
If the eruption process remains ideal as the flux rope is moving
out to infinity, it would have to stretch out its overlying fields
to infinity.  This may be energetically impossible since it has
been shown that for a given normal magnetic flux distribution at the
photosphere, the energy of any three-dimensional force free magnetic field
with all the field lines simply connected to the photosphere is smaller
than the energy of the corresponding fully open magnetic field
(\cite{Aly:1984,Aly:1991,Sturrock:1991}).
This is the so-called Aly-Sturrock energy constraint.
Thus there is not enough energy in a pre-eruption force free field
to stretch all the field lines to infinity to reach a fully open
field.  One way to 
get around this difficulty is that an ideal MHD instability or loss of
equilibrium can extend
the field lines to some height while driving the formation of a current
sheet behind the erupting flux rope, and a fast reconnection in the
driven current sheet allows the stretched-out field lines to successively
reconnect and close back down behind the flux rope as the flux rope
is moving out (e.g. \cite{Priest:Forbes:2002}).
Thus a fast magnetic reconnection is necessary to sustain the
eruption and allow the flux rope to escape to infinity.

\subsection{Sudden onset of fast magnetic reconnection as trigger for eruptions}

Besides the ideal MHD instability and loss of equilibrium processes, another
likely trigger for the onset of eruption is the sudden onset or enhancement of
fast magnetic reconnection in a current sheet.
Detailed theoretical and computational studies of magnetic reconnection have
demonstrated that such sudden onset behavior of fast magnetic
reconnection can be achieved through ``collisionless'' effects when the
current sheet thickness falls to the order of the ion inertial length
(e.g. \cite{Bhattacharjee:2004, Cassak:etal:2005}), or even in the resistive
MHD regime due to the onset of the plasmoid instability for extended thin
current sheets of high Lundquist number (e.g. \cite{Bhattacharjee:etal:2009}).
In the following example MHD simulations that focus on studying the
large scale dynamic evolution of eruptions triggered by the onset of fast
magnetic reconnection, the reconnection processes are modeled with
resistive MHD using a variable magnetic diffusion $\eta$ or a scale
dependent numerical diffusivity.

\begin{figure}[htb!]
\centering
\includegraphics[width=0.99\textwidth]{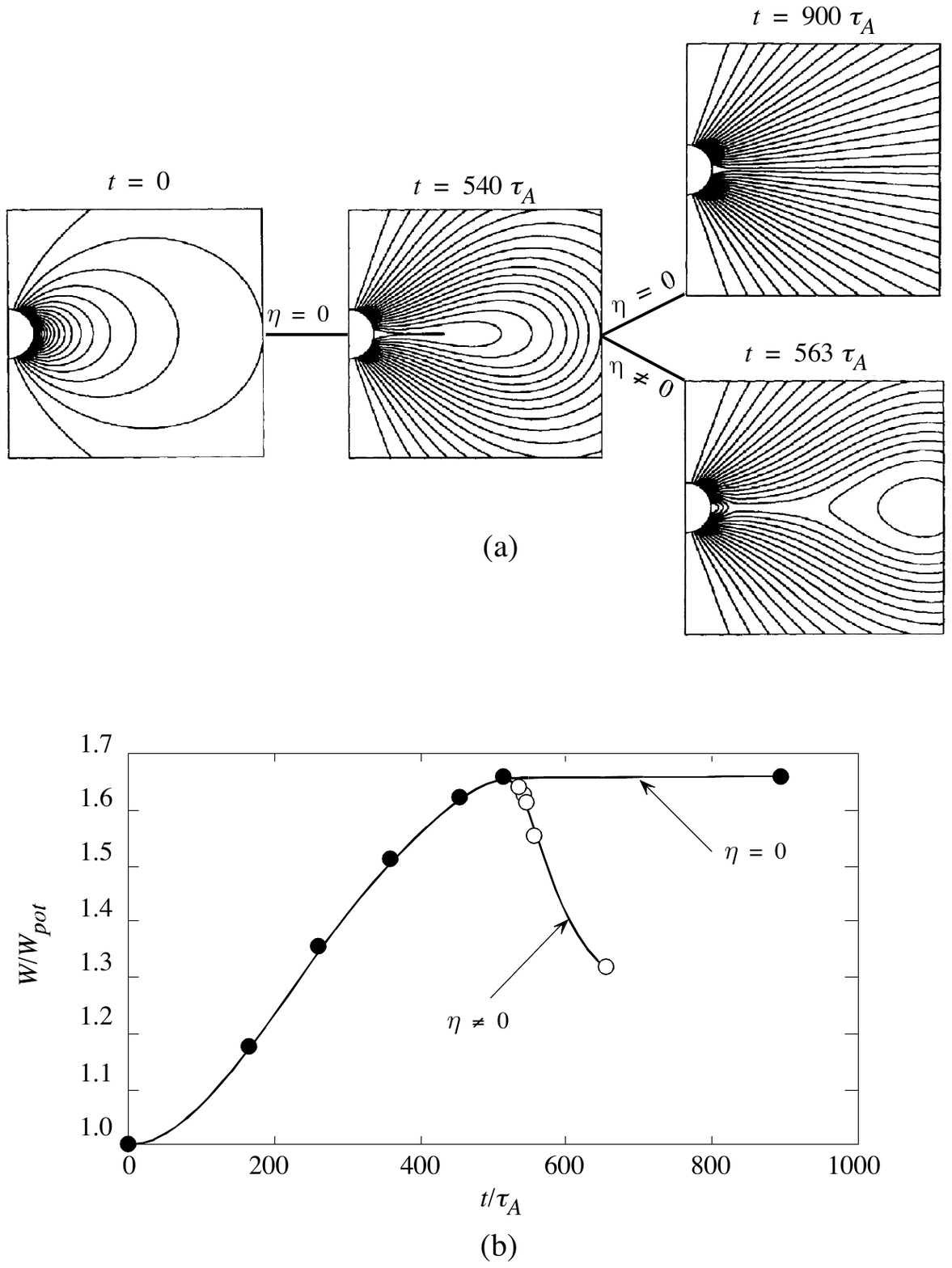}
\caption{Simulation from \cite{Mikic:Linker:1994} of an axisymmetric
coronal arcade field subject to slow shearing motion at its
footpoints on the solar surface with the two hemispheres moving in
opposite directions. The panels in (a) show snapshots
of the magnetic field evolution with $\eta$ remaining zero .vs. being
instantaneously switched on after $t=540 \tau_A$. Panel (b) shows the
evolution of the magnetic energy for the two different cases.
Figure reproduced from the review \cite{Forbes:2000} with permission.}
\label{fig:forbes2000}
\end{figure}

One of the early simulations that clearly demonstrated resistivity triggered
eruption is \cite{Mikic:Linker:1994}, in which an axisymmetric dipolar coronal
arcade field outside the solar surface is subject to a slow shearing motion at
its footpoints (see Figure \ref{fig:forbes2000}).
The initial magnetic field is a Sun-centered dipole field ($t=0$ panel),
which inflates quasi-statically through a sequence of force free
equilibria ($t=540 \tau_A$ and $t=900 \tau_A$ panels, where $\tau_A$ is
the Alfv\'en time defined as the solar radius divided by the Alfv\'en speed)
with increasing magnetic energy
asymptotically towards a fully open field of the maximum energy (solid
black dots in panel (b)) if the evolution remains ideal with the
magnetic diffusivity $\eta = 0$.
For such two-dimensional force free configurations with translational
symmetry, shearing causes the arcade to expand quasi-statically outwards
towards a fully opened state without developing an ideal loss of equilibrium.
However, it is found that during the quasi-static expansion if
$\eta$ is instantaneously increased to a value which gives an
effective magnetic Reynolds number of about $10^4$, rapid reconnection
takes place at the current sheet at the equator, leading to the formation
of a plasmoid which is ejected dynamically (see the $t=563 \tau_A$ panel),
and a sharp release of the magnetic energy (see the white circles in panel (b)).

\begin{figure}[htb!]
\centering
\includegraphics[width=0.99\textwidth]{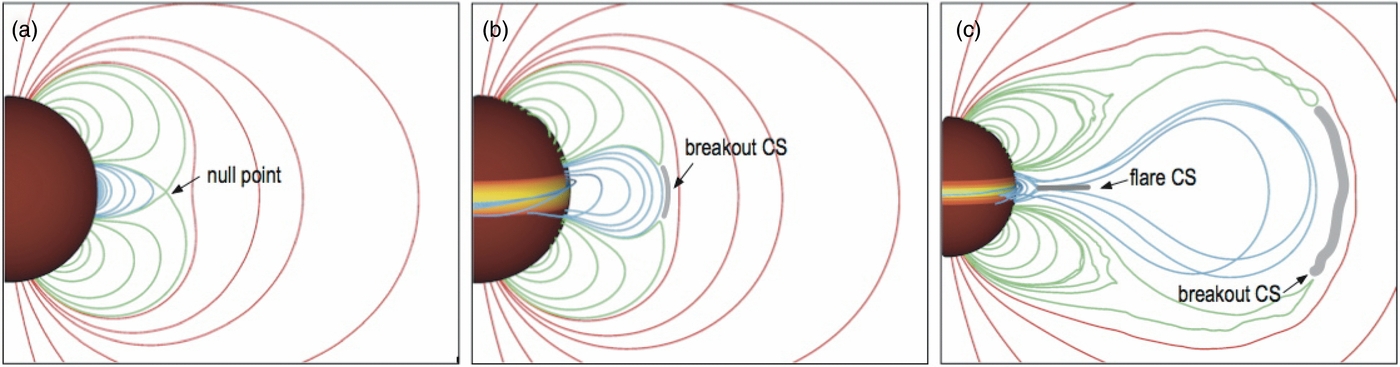}
\caption{Key steps in the simulation of \cite{Karpen:etal:2012} illustrating
the basic working and structure (including current sheets CS) in the break
out model. Figure reproduced from \cite{Karpen:etal:2012} with permission from
the AAS.}
\label{fig:karpenetal2012_fig1}
\end{figure}

\begin{figure}[htb!]
\centering
\includegraphics[width=0.99\textwidth]{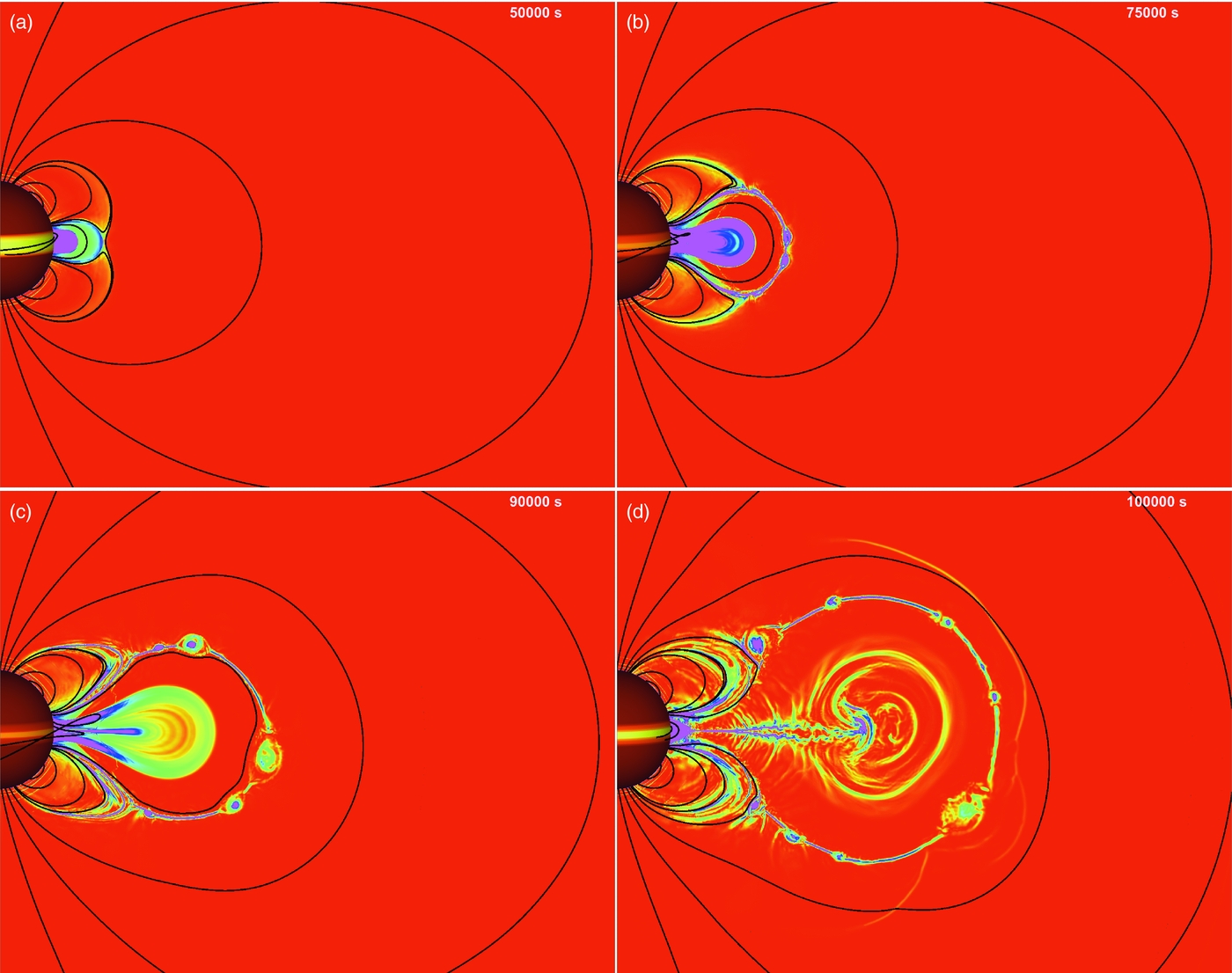}
\caption{$B_{\phi}$ on the solar surface, and normalized current density and
magnetic field lines in the meridional cross-section, at selected times
in the breakout simulation of \cite{Karpen:etal:2012}. Figure from
\cite{Karpen:etal:2012} reproduced by permission of the AAS.}.
\label{fig:karpenetal2012_fig3}
\end{figure}

Another well studied model for CME initiation triggered by magnetic
reconnection is the ``breakout'' model first described by
\cite{Antiochos:etal:1999}.
MHD simulations based on the breakout model have been carried out in both 
2.5D and 3D configurations (e.g. \cite{MacNeice:etal:2004,Lynch:etal:2008}),
and in a 3D configuration with an ambient solar wind
(e.g. \cite{vanDerHolst:etal:2009}).
\cite{Karpen:etal:2012} carried out 2.5D simulations of the breakout model
with adaptive mesh refinement, achieving high resolutions in the current
sheet in global scale dynamic simulations of eruption, and thus allowing
identification of the various key phases of the reconnections in relation
to the eruption.
The basics of how the breakout model works is described as follows (see
Figure \ref{fig:karpenetal2012_fig1}).
The initial potential magnetic field outside the solar surface contains a
multiflux system with three neutral lines on the surface and four distinct
flux systems (Figure \ref{fig:karpenetal2012_fig1}(a)):
a central arcade straddling
the equator (blue field lines), two side arcades associated with the neutral
lines at mid latitudes (green field lines), and a polar flux system overlying
the three arcades (red lines). There are two separatrix surfaces that define
the boundaries between the various flux systems, and a null point above the
central arcade.
In the initial quasi-static phase, the inner arcade field of the central
arcade system is being sheared slowly to become the filament channel (sheared
field) and remains confined by the unsheared arcade to build up free
magnetic energy.  As long as the reconnection between the unsheared 
(blue) arcade of the central system and the overlying red arcade is
slow compared to the shearing, i.e. nearly ideal, energy is built up.
However the current sheet formed at the null point eventually becomes thin
enough (see the breakout CS in Figure \ref{fig:karpenetal2012_fig1}(b)) and
the scale-dependent numerical resistivity causes a sudden onset of the fast
magnetic reconnection in the breakout CS (see Figure
\ref{fig:karpenetal2012_fig3}(b)).
The fast breakout reconnection
rapidly transfers the confining flux in the unsheared arcade and
the overlying outer arcade into the side lobes, causing an outward
expansion of the filament channel field and the confinement is found to
be permanently lost. Thus the onset of the breakout reconnection is the
trigger of the CME and if it were to continue to act alone
(Figure \ref{fig:karpenetal2012_fig3}(c)), it would
have led to a complete but slow outward expansion of the filament channel
field with a vertical current sheet extending downward to low heights in
the corona, and thus with most of the free energy retained. But then a
sudden onset of fast reconnection in the flare current sheet is found
to trigger impulsive energy release with the formation of a flux rope
that accelerates impulsively (Figure \ref{fig:karpenetal2012_fig1}(c) and
\ref{fig:karpenetal2012_fig3}(d)).
Thus in this model, the resistive process of the onset of fast
magnetic reconnection is the initiation mechanism (or trigger) for both
the CME onset (loss of equilibrium for the filament channel field) as well
as the later onset of impulsive CME acceleration.
For the former it is the onset of the fast breakout reconnection and for the
latter it is the onset of fast reconnection in the flare CS.
The question then is whether it is possible to observationally identify these
two stages of the development.

\section{Concluding remarks}
The question of how twisted and sheared coronal magnetic structures
capable of driving filament eruptions form is examined with focus on emerging
active regions.  MHD simulations of flux emergence have shown that many
commonly observed features associated with flare and CME productive active
regions, such as magnetic and velocity shear at the PILs,
rotating sunspots, sigmoid-shaped X-ray loops and filaments, can be explained
by the emergence of a twisted magnetic flux tube from the solar interior into
the atmosphere.
It is found that shear and twisting motions driven by the Lorentz force
of the emerging tube are the major means (instead of direct vertical
emergence through the photosphere) by which magnetic helicity (twist) and
energy are transported into the corona.
It is also found that current sheet formation and the associated
tether-cutting reconnections are important for the buildup of the coronal
flux rope and allow it to rise into the corona. This is found in both
simulations of flux emergence through the photosphere as well as simulations
of line-tied coronal flux ropes evolving quasi-statically towards the onset
of eruption.

Several basic mechanisms that can trigger the sudden disruption of the
quasi-equilibrium coronal magnetic structures and explosive release of
the stored free magnetic energy are discussed.  These include both
ideal processes such as the onset of the helical kink instability and the
torus instability of a coronal flux rope, and the non-ideal processes of
the onset of fast magnetic reconnections in a current sheet.
It is found that even when the triggering mechanism is an ideal instability or
loss of equilibrium of a coronal flux rope, magnetic reconnection is playing
an important role in its buildup during the quasi-static stage, as well as for
sustaining the eruption and allowing for an ejection of flux rope due to
the Aly-Sturrock energy constraint.
MHD simulations have shown that such ``hybrid'' models of the quasi-static
buildup of torus unstable coronal flux ropes with the development of the ``HFT''
topology can explain some of the thermal features observed to develop in
pre-eruption coronal prominence cavities.  MHD simulations of the non-linear
evolution of kink-unstable coronal flux ropes have also shown magnetic field
evolution that resembles remarkably well the observed morphology of some of
the highly writhing filament eruptions.
Fast magnetic reconnection can also trigger eruptions as described in 
the breakout model without the presence of a more twisted pre-eruption
coronal flux rope capable of developing ideal instabilities.
Simulations based on this model show that the onset of fast reconnection in
the flare current sheet that rapidly creates a highly twisted flux rope is
needed for the onset of the impulsive acceleration of the CMEs.

It should be noted that the discussion of this chapter has focused entirely
on nearly force free coronal structures, ignoring the non force free effects
of the prominence weight for example.  There is an important body of
work on the role of prominence mass in energy storage of
the prominence magnetic field and CME energetics (e.g. \cite{Low:Smith:1993,
Low:Zhang:2002,Fong:Low:Fan:2002,Low:Fong:Fan:2003,
Zhang:Low:2004,Zhang:Low:2005}), especially with regard
to overcoming the Aly-Sturrock energy constraint.

\begin{acknowledgement}
The author would like to thank Fang Fang, Terry Forbes, Judy Karpen,
Zoran Mikic, Slava Titov, and Tibor T{\"o}r{\"o}k for granting permission to
reprint Figures from their original publications.
The National Center for Atmospheric Research is sponsored by the National
Science Foundation.
\end{acknowledgement}

\newpage
% BibTeX users please use
%\bibliographystyle{spmpsci.bst}
%\bibliography{references}
%
% Non-BibTeX users please follow the syntax
% the syntax of "referenc.tex" for your own citations
%\input{referenc}
%%%%%%%%%%%%%%%%%%%%%%%%%%%%%%%%%%%%%%%%%%%%%%%%%%%%%%%%%%%%%%%%%%%%%%

%%%%%%%%%%%%%%%%%%%%%%%%%%%%%%%%%%%%%%%%%%%%%%%%%%%%%%%%%%%%%%%%%%%%%%

\printindex
\end{document}